\documentclass[prl,twocolumn,superscriptaddress,showpacs]{revtex4}
\usepackage{amsmath,amssymb}
\usepackage{graphicx}

\begin{document}

\title{Bloch-wave engineering of quantum dot-micropillars for cavity quantum electrodynamics experiments}

\author{M. Lermer}
 \affiliation{Technische Physik, Wilhelm Conrad R\"{o}ntgen Research Center for Complex Material Systems, Universit\"{a}t W\"{u}rzburg, Am Hubland, D-97074 W\"{u}rzburg, Germany}

\author{N. Gregersen}
 \affiliation{DTU Fotonik, Department of Photonics Engineering, Technical University of Denmark, Building 343, DK-2800 Kongens Lyngby, Denmark}

\author{F. Dunzer}
 \affiliation{Technische Physik, Wilhelm Conrad R\"{o}ntgen Research Center for Complex Material Systems, Universit\"{a}t W\"{u}rzburg, Am Hubland, D-97074 W\"{u}rzburg, Germany}

\author{S. Reitzenstein}
 \affiliation{Technische Physik, Wilhelm Conrad R\"{o}ntgen Research Center for Complex Material Systems, Universit\"{a}t W\"{u}rzburg, Am Hubland, D-97074 W\"{u}rzburg, Germany}

\author{S. H\"{o}fling}
 \affiliation{Technische Physik, Wilhelm Conrad R\"{o}ntgen Research Center for Complex Material Systems, Universit\"{a}t W\"{u}rzburg, Am Hubland, D-97074 W\"{u}rzburg, Germany}

\author{J. M{\o}rk}
 \affiliation{DTU Fotonik, Department of Photonics Engineering, Technical University of Denmark, Building 343, DK-2800 Kongens Lyngby, Denmark}

\author{L. Worschech}
 \affiliation{Technische Physik, Wilhelm Conrad R\"{o}ntgen Research Center for Complex Material Systems, Universit\"{a}t W\"{u}rzburg, Am Hubland, D-97074 W\"{u}rzburg, Germany}

\author{M. Kamp}
 \affiliation{Technische Physik, Wilhelm Conrad R\"{o}ntgen Research Center for Complex Material Systems, Universit\"{a}t W\"{u}rzburg, Am Hubland, D-97074 W\"{u}rzburg, Germany}

\author{A. Forchel}
 \affiliation{Technische Physik, Wilhelm Conrad R\"{o}ntgen Research Center for Complex Material Systems, Universit\"{a}t W\"{u}rzburg, Am Hubland, D-97074 W\"{u}rzburg, Germany}

 \date{today}

\begin{abstract}
We have employed Bloch-wave engineering to realize submicron diameter ultra-high quality factor GaAs/AlAs micropillars (MPs).  The design features a tapered cavity in which the fundamental Bloch mode is subject to an adiabatic transition to match the Bragg mirror Bloch mode. The resulting reduced scattering loss leads to record-high visibility of the strong coupling in MPs with modest oscillator strength quantum dots. A quality factor of 13,600 and a Rabi splitting of $85~\mu$eV with an estimated visibility $v$ 
of $0.38$ are observed for a small mode volume MP with a diameter $d_c$ of $850~$nm.
\end{abstract}


\maketitle

Semiconductor quantum dot (QD) - microcavity systems are highly interesting for the study of fundamental light-matter interaction and the development of innovative quantum devices \cite{Gerard2003}. As the strength of the light-matter interaction depends strongly on the electromagnetic field profile, it is central to engineer the photonic environment in these systems to obtain the desired functionality. In devices featuring invariance along a propagation z-axis, the modal formalism \cite{Snyder1983} gives a clear physical picture of the governing mechanisms of the propagation and scattering of light, which allows for a simple understanding of the effects of e.g. adiabatic mode transitions, mode coupling and the scattering of a truncated waveguide. Thanks to the Bloch theorem, a similar modal formalism exists for periodic structures, and most of the physical concepts used for z-invariant geometries can be carried over to periodic structures. The potential of Bloch-wave engineering in microcavities was pointed out by Lalanne et al. \cite{BlochWaveTheo} and has subsequently been experimentally demonstrated in planar 1D photonic crystal (PhC) nanobeam cavities \cite{BlochWaveExp,Ohta2011}
and in 2D PhC cavities \cite{Song2005}. Furthermore, 'accidental' improvements in cavity figure of merits \cite{Akahane2003} have subsequently been explained using Bloch-wave concepts \cite{Sauvan2004}. However, in spite of the clear potential for investigating light-matter interaction in Bloch-wave engineered MP cavities \cite{Zhang2009}, which have previously been utilized as a model system for the demonstration of cavity quantum electrodynamics (cQED) effects such as strong light-matter coupling \cite{Reithmaier2004},  efficient emission of single \cite{SPS} and indistinguishable \cite{Santori2002} photons and entangled photon pairs \cite{Dousse2010},  no experimental demonstration of intentionally engineered MP cavities has been reported so far.

\begin{figure}
\begin{center}
\includegraphics[angle=0,width=\linewidth]{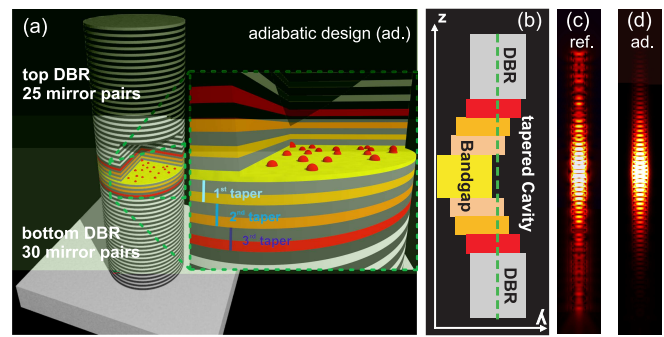}
\caption{(a) Schematic sketch of the layer sequence of the adiabatic cavity design illustrating the bottom/top DBR with 30/25 mirror pairs and a 3 taper cavity region. Inset: Zoom-in view of the cavity region. (b) The position of the photonic bandgap as function of z is schematically illustrated. The resonance wavelength of the cavity mode is represented by the dashed line.  (c, d) Electric field amplitude profiles in a MP with $d_c = 750~$nm for the reference (c) and the adiabatic (d) design, respectively. }
\end{center}
\end{figure}

In this Letter, we present a novel tapered MP cavity design in which Bloch-wave engineering is employed to significantly enhance the 3D mode confinement and thereby the light-matter interaction. For the observation of the Purcell-effect in a weakly coupled QD-cavity system, the figure of merit is the ratio $Q/V_{mode}$ \cite{Gerard2003}, where $Q$ is the quality factor (Q factor) and $V_{mode}$ is the mode volume, and in the strong coupling (SC) regime the normal mode splitting is resolved when the visibility  $v=g/(\kappa +\gamma^*)>1/4$  \cite{Auffeves2010}. Herein, $\gamma^*$ is the pure dephasing rate, $\kappa = \omega_c/Q$ represents the full width at half maximum of the fundamental mode with resonance frequency $\omega_c$ and the slow decay rate of the isolated emitter is neglected. The light-matter coupling constant $g$ is given by $g=\sqrt(e^2f/(4\varepsilon_0 n_c^2 m_0 V_{mode}))\propto 1/d_c$, where $f$ is the oscillator strength and $m_0$ is the free-electron mass \cite{Gayral2008}. In both regimes it is thus necessary to confine light in microresonators providing high Q factors and small mode volumes.
MP cavities have been demonstrated to support photonic modes with quality factors as large as 165,000 \cite{Reitzenstein2007} in the few $\mu$m diameter ($d_c$) range and thus relatively large $V_{mode}$. To observe SC, a large $V_{mode}$ can generally be compensated by increasing Q, however, when $\kappa <\gamma^*$, the pure dephasing rate limits the visibility. This important effect means that increasing Q beyond  $\omega_c/\gamma^*$ does not improve $v$, which in the few $\mu$m diameter range generally remains $\leq1/4$ for standard QDs \cite{Reitzenstein2007}.

Rather, in the presence of dephasing, the figure of merit for SC becomes $g/(\kappa+\gamma^*)$ \cite{Auffeves2010} and the visibility is maximized by choosing a high Q on the order of $\omega_c/\gamma^*$ and by reducing $V_{mode}$. To this end, one might simply decrease $d_c$ of the pillars. However, in the standard microcavity design, the mode matching between the cavity mode and the fundamental Bloch mode of the distributed Bragg reflectors (DBRs) in the submicron regime is poor, leading to intrinsic scattering loss and a Q factor on the order of 2,000 for $d_c < 1\mu$m \cite{SubmicronMC}. Moreover, rather than being a monotonous function of diameter, coupling of the cavity mode to propagating Bloch modes of the DBRs has been shown to result in substantial Q factor oscillations \cite{SubmicronMC, Reitzenstein2009, Gregersen2010}. Whereas the Q factor of large-diameter MPs with $d_c > 3\mu$m \cite{Reitzenstein2007} is mainly limited by scattering from fabrication imperfections and intrinsic loss, the poor mode matching is the main limitation for submicron diameter pillars.

To reduce the strong scattering due to mode mismatch in the submicron regime we have implemented an advanced AlAs/GaAs pillar design inspired by a proposal of adiabatic SiO$_2$/TiO$_2$ MP cavities by Zhang et al. \cite{Zhang2009}. The design features a tapered cavity region in which the fundamental DBR Bloch mode experiences an adiabatic mode transition \cite{BlochWaveTheo} out of the bandgap. We obtain a strongly improved mode matching and a reduced coupling to propagating Bloch modes in the DBRs, and this allows us to demonstrate remarkably high quality factors exceeding 10,000 for MP cavities with diameters below $1~\mu$m.

Whereas previous studies on MP cavities required large oscillator strength QDs to enter the SC regime \cite{Reithmaier2004, Press07, Kasprzak2010, Loo2010}, fabricated using a highly challenging process, the good mode matching for adiabatic MPs with $d_c < 1\mu$m paves the way for pronounced strong coupling effects even for standard InGaAs QDs with more pronounced quantization effects and oscillator strength $f$ of about 10.

Our advanced approach consists of employing an adiabatic Bloch mode matching technique for the GaAs/AlAs cavity region. A standard design with a regular one-$\lambda$  cavity is also studied as a reference. In the 'adiabatic' design the regular $\lambda$-cavity is replaced by a 3 segment tapered region. The layer thicknesses are gradually reduced towards the center, effectively detuning the bandgap relative to that of the DBRs, allowing for a single localized mode inside the cavity \cite{Zhang2009}. The geometrical parameters were determined using full 3D vectorial simulations based on the eigenmode expansion technique \cite{EET} for a design diameter of 1,000 nm and a resonance wavelength $\lambda_c=2\pi c/\omega_c= 950~$nm. Fabrication imperfections, i.e. the partial oxidation of the AlAs layers and the coating of the pillars with glass (SiO$_x$) during the dry etching step \cite{Reitzenstein2009, Gregersen2010}, were taken into account. The DBR layer thicknesses were chosen as $\lambda_{eff}/4=\lambda_c/(4n_{eff})$, where $n_{eff}$ is the effective index of the layer and $\lambda_{eff}$ the effective material wavelength, placing the design resonance at the center of the 1D DBR bandgap. A simple linear decrease of the cavity layer thicknesses was employed and the linear coefficient was adjusted to obtain a resonance at $\lambda_c = 950~$nm. Fig. 1(a) shows a schematic of the layer sequence of the adiabatic cavity design. On the right hand side a zoom-in of the tapered cavity region is shown, illustrating by different colors the decrease of the thickness of the layers upon transition from the periodic DBR region towards the QD-layer, which is placed in the central GaAs spacer.

Whereas the cavity thickness of the reference design is $\lambda_{eff}=280~$nm, the central GaAs layer in the tapered cavity is only 60 nm thick, corresponding to about $\lambda_{eff}/5$. Regular cavity engineering concepts are thus insufficient to explain the localization of the cavity mode in the adiabatic pillar, which demonstrates the necessity of Bloch mode formalism in the analysis of the geometry. Instead of considering eigenmodes in z-invariant layers we consider Bloch modes in sections of alternating GaAs/AlAs layers. Somewhat surprisingly, Bloch modes remain useful even in regions of the adiabatic cavity containing only one period of layer pairs. Furthermore, the introduction of the concept of Bloch modes in these aperiodic regions serves to fully explain the suppressed scattering due to mode mismatch, as we show in the following. The position of the bandgap in the adiabatic cavity as function of the vertical z coordinate is illustrated in Fig. 1(b). Outside the cavity, the bandgap in the DBRs is centered at the resonance wavelength to achieve maximum reflectivity. Inside the cavity, the total thickness of the GaAs/AlAs layer pairs is gradually decreased. This decrease leads to a gradual blueshift of the bandgap ensuring an adiabatic transition of the fundamental Bloch mode, and in the central region it appears outside the bandgap. The Bloch mode can propagate freely over 2 GaAs/AlAs periods, as illustrated in Fig. 1(b), before it is gradually reflected, resulting in the localization of a cavity mode in the central region combined with suppression of the scattering to higher-order Bloch modes. Optical field profiles of MPs with $d_c = 750~$nm are plotted in Fig. 1(c-d). The field profile for the reference design in Fig. 1(c) depicts a localized mode, but also substantial field penetration into the DBRs due to coupling to propagating Bloch modes. Figure 1(d) shows the cavity mode for the adiabatic cavity design, and here the suppressed coupling to higher-order Bloch modes, the reduced field penetration into the DBRs and the improved confinement are clearly observable.

The computed Q and $V_{mode}$ for the adiabatic and the reference designs are shown in Fig. 2(a). The simulations show diameter dependent oscillatory variations with a small period ($\approx50~$nm) of the Q factor between 100,000 and 300,000 for the adiabatic cavity design (see Fig. 2(a)). This is a theoretical improvement of Q by 2 orders of magnitude for the adiabatic cavity design compared to the reference band results from the strong improvement of mode matching between the cavity and DBR Bloch modes. Even though the design was optimized for $d_c = 1,000~$nm, we observe that the Q factor remains large in a broad diameter range. Fig. 2(b) shows that the Q factor of MP cavities based on the reference design is not limited by the amount of mirror pairs for $d_c < 1,000 ~$nm, confirming that the mode mismatch is the dominating limiting effect on the Q factor in this diameter regime. One also recognizes from Fig. 2(a) that $V_{mode}$ of the two designs are almost identical, leading in total to a theoretical enhancement of Q/$V_{mode}$ by 2 orders of magnitude for the adiabatic cavity design compared with the standard $\lambda$-thick cavity.

\begin{figure}
\begin{center}
\includegraphics[angle=0,width=\linewidth]{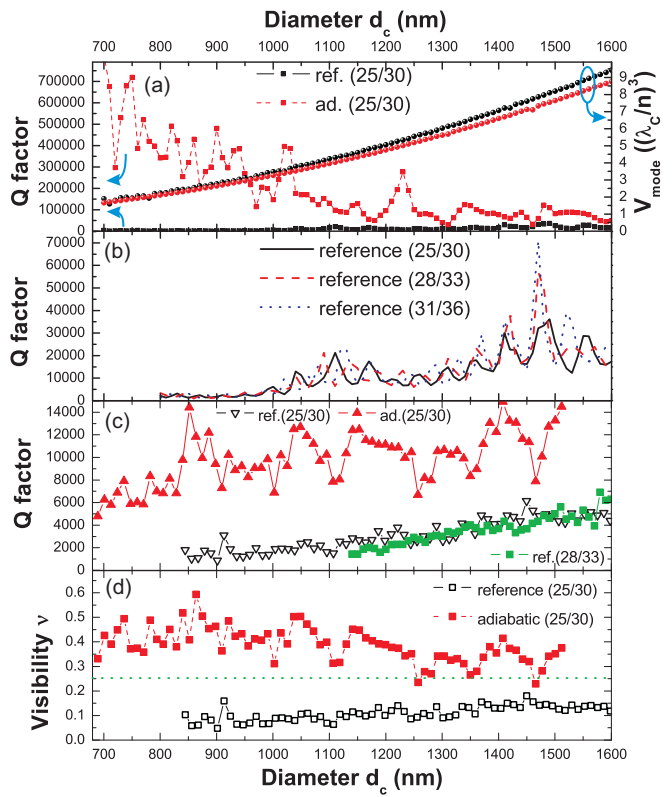}
\caption{(a) Computed Q factor and $V_{mode}$ as function of $d_c$ for the reference and the adiabatic cavity designs. (b) Computed Q factor for reference designs with 25/30, 28/33 and 31/36 DBR mirror layer pairs. (c) Experimentally measured maximum values of the Q factor for the reference (25/30 and 28/33 mirror pairs) and the adiabatic cavity designs (25/30 mirror pairs). (d) The visibility as function of diameter $d_c$ is illustrated for the adiabatic (dashed curve) and reference design (full curve) for an oscillator strength of $f = 10$. The onset of SC at $v= 1/4$ is depicted using the dotted line.}
\end{center}
\end{figure}

The AlAs/GaAs microcavity samples have been grown by molecular beam epitaxy on (100) oriented GaAs substrate. Each sample consists of a cavity region, sandwiched between an upper (lower) distributed Bragg reflector with 25 (30) and 28 (33) mirror pairs, composed of alternating AlAs and GaAs layers. MP cavities with various diameters have been fabricated by electron beam lithography and dry etching techniques and characterized by  micro-photoluminescence spectroscopy under above band gap excitation at $532~$nm.

Fig. 2(c) shows the measured Q factor dependent on $d_c$ for the reference and adiabatic cavity designs. One observes that of the reference design is limited to values of about 2,000 in the submicron diameter range, which is in agreement with literature \cite{SubmicronMC}. Similar Q factors are observed for both reference samples with 25/30 and 28/33 DBRs, consistent with the simulations shown in Fig. 2(b). Significantly higher Q factors are obtained for the MPs based on the adiabatic cavity design. Indeed, all Q values of the improved microcavity design are located in the range between 4,000 and 15,000 in the submicron diameter range. Interestingly, a Q factor as high as 14,400 has been identified for a MP cavity with $d_c = 860$nm which is a factor of 7 higher than state-of-the-art values for submicrometer MPs with standard microcavity design \cite{SubmicronMC}.

Simulations predict an improvement in Q of two orders of magnitude and predict a general increase in Q as $d_c$ decreases, whereas the opposite trend is observed experimentally. We attribute these discrepancies to random fabrication-induced imperfections not taken into account in the simulations. For the perfectly periodic structures considered in the modeling, the calculations show that the poor modal overlap in the submicron diameter regime is indeed the dominating mechanism limiting the Q factor and the adiabatic transition of the fundamental Bloch mode leads to a vast improvement in Q. However, random side-wall imperfections for the experimentally realized pillars are present resulting in a scattering loss inversely proportional to the pillar diameter. While this loss is less important for large-diameter pillars \cite{Reitzenstein2007}, scattering by side-wall imperfections is the dominating loss mechanism for $d_c < 1,000~$nm and limits the obtainable values of Q in this regime.

\begin{figure}
\begin{center}
\includegraphics[angle=0,width=\linewidth]{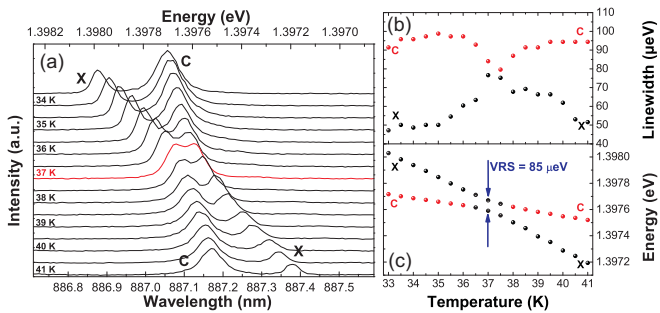}
\caption{(a) Strong QD (X) - cavity (C) coupling of a MP with $d_c = 850~$nm and $Q ~ 13,600$ via temperature tuning. Temperature dependent (b) linewidth and (c) peak position determined from (a).}
\end{center}
\end{figure}

The adiabatic microcavity design is of particular interest for cQED experiments in the SC regime. Due to the combination of a small $V_{mode}$ and a high Q factor, the observation of normal mode splitting in submicron pillars with 'standard' QDs having oscillator strengths $f \sim 10$ should be feasible. For the calculation of the visibility v, we estimate a dephasing rate $\gamma^*$ of $ \sim 20~\mu$eV at $35~$K \cite{Loo2010, LAphonons}. The resulting visibility $v$ as function of $d_c$ for the adiabatic and the reference designs obtained from the measured Q factor and the computed $V_{mode}$ for a standard oscillator strength of $f = 10$. The visibility of the reference design falls much below the onset for SC at $1/4$. However, the visibility for the adiabatic cavity design increases well above $1/4$, and the observation of SC can be expected. Indeed, we have identified SC for various submicrometer diameter QD-MPs based on the adiabatic cavity design. One example is presented in Fig. 3, where temperature tuning of a single QD-line (X) with the fundamental mode (C) of a MP with $Q = 13,600$ and $d_c = 850$ nm ($V_{mode} \approx 2.3\, (\lambda_c/n^3$)) is applied to demonstrate anticrossing of the two lines (see Fig. 3 (a)) as a clear signature of the coherent coupling. At resonance (T = 37 K), X and C are indistinguishable in terms of their widths (see Fig. 3 (b) and (c)) and a Rabi splitting of $85~\mu$eV, can be extracted, corresponding to a minimum QD oscillator strength of $f \approx 3$  and a visibility $v$ of $0.38$, to our knowledge representing state-of-the-art in micropillars with standard ($f \approx 10$) QDs.

In conclusion, we have employed Bloch-wave engineering to realize ultra high Q factor MP cavities in the sub-micrometer diameter range. Instead of the usual $\lambda$-cavity the design features a symmetric 3 segment tapered cavity where adiabatic transition is employed to reduce the usual scattering due to poor mode matching occuring in standard submicrometer diameter pillars. The adiabatic cavity design has lead to experimentally measured Q factors exceeding 10,000 for pillar diameters $d_c < 1,000$nm. This improvement makes the design very promising for the observation of pronounced cQED effects, and SC between a single QD line and the fundamental mode of a submicron diamater MP has been observed. While enormous effort has been invested in tailoring 2D and 3D photonic crystal cavities with small $V_{mode}$ and high Q factors, Bloch-wave engineering of MP cavity designs has so far been largely neglected. The present study shows the large potential of employing Bloch-wave engineering for solid-state cQED.

\begin{acknowledgments}
 We thank A. Wolf and M. Emmerling for sample preparation and P. K. Nielsen and K. H. Madsen for fruitful discussions.  Financial support by the Federal Ministry of Education of Research (project 'QuaHL-Rep') and the Danish Research Council for Technology and Production (Contract No. 10-080752) is gratefully acknowledged.\end{acknowledgments}

\end{document}